# Experimental manifestations of the $Nb^{4+}$-$O^-$ polaronic excitons in $KTa_{0.988}Nb_{0.012}O_3$

R.V. Yusupov[1], I.N. Gracheva[1], A.A. Rodionov[1], P.P. Syrnikov[2], A.I. Gubaev[3], A. Dejneka[4], L. Jastrabik[4], V.A. Trepakov[2,4], and M.Kh. Salakhov[1]

[1] Kazan (Volga Region) Federal University, 420008, Kremlevskaya 18, Kazan, Russia
[2] Ioffe Physical-Technical Institute RAS, 194021 St.-Petersburg, Russia
[3] Institute for Physical Chemistry, Münster University, Corrensstrasse 30D-48149, Germany
[4] Institute of Physics AS CR, Na Slovance 2, 182 21 Prague 8, Czech Republic

## Abstract

The formation of the photo-polaronic excitons in $ABO_3$ perovskite type oxides has been detected experimentally by means of the photoinduced electron paramagnetic resonance studies of $KTa_{0.998}Nb_{0.012}O_3$ crystals. The corresponding microwave X-band spectrum at $T < 10$ K consists of a narrow, nearly isotropic signal located at $g \sim 2$ and a strongly anisotropic component. The first signal, which has a rich structure due to hyperfine interactions with the lattice nuclei, is attributed to the single trapped charge carriers: the electrons and/or the holes. The anisotropic spectrum is caused by the axial centers oriented along the $C_4$ pseudo-cubic principal crystalline axes. The spectrum angular dependence can be described well by an axial center with $S = 1$, $g_\parallel = 0.82$, $g_\perp = 0.52$ and $D = 0.44$ cm$^{-1}$. The anisotropic spectrum is attributed to the $Nb^{4+}$-$O^-$ polaronic excitons. The temperature dependence of the anisotropic component is characterized by two activation energies: the internal dynamics activation $E_{a1} = 3.7\pm0.5$ meV, which makes the EPR spectrum unobservable above 10 K, and the destruction energy $E_{a2} = 52\pm4$ meV. By comparing the anisotropic photo-EPR spectrum and the photoinduced optical absorption temperature dependencies, we found that the $Nb^{4+}$-$O^-$ polaronic excitons also manifested themselves via the ~0.7 eV wide absorption band arising under UV light excitation in the weakly concentrated $KTaO_3$:Nb crystals.

## INTRODUCTION

Currently, there are many studies of systems near the quantum mechanical limit, i.e., when the systems experience phase transition at zero or extremely low temperatures. Some specific properties of the heavy-fermion systems and the high-temperature superconductors are associated with the proximity to the *"quantum critical point"*, i.e., a zero-temperature transition. In the low-temperature region, the ground quantum state of such systems experiences drastic modifications. Due to the inherent relation between the static and the dynamic properties of the quantum systems, temporal characteristics strongly influence the system properties in the critical region.

The "quantum paraelectrics" $KTaO_3$ (KTO) and $SrTiO_3$ (STO) are the most intensely studied representatives of the highly polarizable $ABO_3$ perovskite-like oxides, which can be considered as systems that are close to the quantum mechanical limit (QML) at low temperatures. In such quantum systems, unlike in the classical case, the transition can be achieved by tuning not the temperature but the other parameters, such as pressure, chemical composition and magnetic field. In this respect, the impurity-induced ferroelectric (FE) phase transitions (PT) in KTO serve as the popular models to study the impurity doping effects in the functional $ABO_3$ perovskite-type oxide family. The giant photodielectric effect recognized recently in STO and KTO (e.g., [1-5]) is another example, where UV light irradiation at low temperatures strongly enhances the dielectric constant and the persistent photoconductivity. Qualitatively, these phenomena have been assigned to the formation of an inhomogeneous polar state induced by the photo-generated charge carriers, but little understanding of the microscopic nature of the polar state and the photo-carriers has been achieved yet.

To solve this problem, we have studied the photoinduced (PI) optical absorption, photoconductivity and luminescence of $KTa_{1-x}Nb_xO_3$ (KTN) single crystals with $0 \leq x \leq 0.07$ [6, 7]. There are two reasons to choose this material. First, the actual photodielectric effects are strongly enhanced as the temperature approaches 0 K, i.e., the critical point [2, 3, 5]. Second, all researchers emphasized the important role of the photo-polarons in the photodielectric effect formation, although the presence of polaronic states had not been strictly verified. Furthermore, this material is promising for observation of the $Nb^{4+}$ polaron formation; also, by tuning the Nb concentration in KTN, one can access the desirable ordered state or the near-QML state. Thus, $x = 0$ corresponds to the nominally pure KTO, which possesses the cubic $O_h^1$ perovskite structure at least down to 80 mK [8]. Its low frequency resonance-type dielectric constant rises on cooling due to TO1 mode softening. However, at low temperatures in the range of dominating quantum statistics, permittivity saturates due to the contribution of quantum fluctuations to the generalized force constant corresponding to the soft FE polar TO1 mode. Therefore, KTO is conventionally called a "quantum paraelectric" [9]. However, KTO is rather an "incipient displacive ferroelectric" or a "soft incipient displacive ferroelectric" because the inequality $|k_h| \ll k_0$ holds, where $k_h$ is the harmonic force constant corresponding to the soft ferroelectric (polar) mode, and $k_0 = k_h + k_{zp}$, where $k_{zp}$ is a contribution from the zero-point vibrations [10]. For $x = 0.008$, $KTa_{1-x}Nb_xO_3$ exhibits the properties of a ferroelectric in the quantum limit [11]: the critical exponent value is 2 for $\varepsilon'(T)$ (instead of the classical 1), all responses are temperature independent near $T = 0$ K and the increase of the niobium concentration leads to the appearance of the ferroelectric PT. For example, $KTa_{0.998}Nb_{0.012}O_3$ (further denoted as KTN-1.2) is a low-T ferroelectric close to QML that obeys a cubic-trigonal PT at $T_C = 16$ K [12]. As a result, our earlier photochromic investigations of diluted KTN crystals, supplemented by photoconductivity and luminescence studies [6, 7], showed a strong localization of the photocharge carriers on the Nb – Nb pairs at low temperatures. This localization can play a crucial role in the aforementioned photoinduced dielectric effects and photoinduced phase transitions.

In this work, we study the X-band photoinduced electron paramagnetic resonance (photo-EPR) of the KTN-1.2 crystals as a logical continuation of our previous work [6, 7]. The anisotropic component of the photo-EPR spectrum was found with $g_{eff\|} = 2.106$ and $g_{eff\perp} < 0.8$, revealing the presence of the axial centers oriented along the $C_4$ principal crystalline pseudo-cubic axes. The angular dependence of the anisotropic component resonance field cannot be described by the $S = \frac{1}{2}$ model with the axial $g$-factor used to describe the photo-EPR spectra in KTO. Meanwhile, a model of the center with $S = 1$, axial $g$-factor and zero-field splitting $D$ slightly larger than the microwave quantum gives a good description of the angular variation of this component resonance field. The temperature variation of the anisotropic spectrum intensity indicates that this spectrum is caused by the same microscopic object responsible for the 0.7 eV photoinduced optical absorption band. Finally, we found the $Nb^{4+}$ photo-polaron formation proposed in [6, 7] for the treatment of the photochromic and photoconductivity data did not reflect the full complexity of the situation. Our new findings indicate that UV illumination of the KTN-1.2 leads to the formation of a more sophisticated object – the $Nb^{4+}$- $O^-$ polaronic exciton.

**EXPERIMENTAL DETAILS**

The single crystals of $KTa_{1-x}Nb_xO_3$ employed in the present EPR investigations were grown by solidification from a non-stoichiometric melt of the ultrapure (99.999 %) $Ta_2O_5$, $K_2CO_3$ and $Nb_2O_5$ starting materials. The KTN-1.2 composition was chosen as a weakly Nb concentrated system close to QML, in which the intense PI optical absorption in the near-IR range assigned to the photo-induced polarons was observed [6, 7]. The specimens were oriented along the cleaved (001) cubic planes. EPR spectra were registered using the commercial continuous wave (*cw*) X-band Bruker ESP300 spectrometer equipped with a standard $TEM_{012}$ rectangular cavity with $B_1 \perp B_0$ ($B_1$ is the magnetic component of the microwave and $B_0$ is an applied static magnetic field). The temperature of the sample was controlled with the commercial Oxford Instruments ESR 9 continuous flow cryogenic system. The specimen was attached with its cleaved (001) plane to the end of the fused silica rod, which, assembled with an optical fiber, served as a waveguide for the UV illumination.

The rotation about the rod axis allowed us to study the EPR spectrum angular dependencies in the $C_4$-$C_2$-$C_4$, (001), crystal plane. The UV radiation source was a Philips XBO high-pressure xenon arc lamp. The spectral selection was performed with a dichroic mirror reflecting the light with wavelengths shorter than 450 nm into the waveguide, allowing the near band-gap and interband excitations.

## RESULTS

Fig. 1 presents the EPR spectra of the KTN-1.2 crystals measured at 4.0 K while keeping the sample in the dark and while exposing it to a stationary UV source. The weak intensity lines with $g_{eff} > 2$ in both spectra are due to the presence of a small amount of iron ions in the cavity but not in the sample and will not be considered here. The photo-EPR spectrum emerging under UV illumination consists of two kinds of signals. The first one (Signal I) is a relatively narrow signal with $g_{eff} \approx 2$. Such signal was also found in the photo-EPR spectra of the KTO crystal. The second photo-EPR signal (Signal II) is represented by a wide asymmetric band with a maximum microwave absorption at ~ 500 mT.

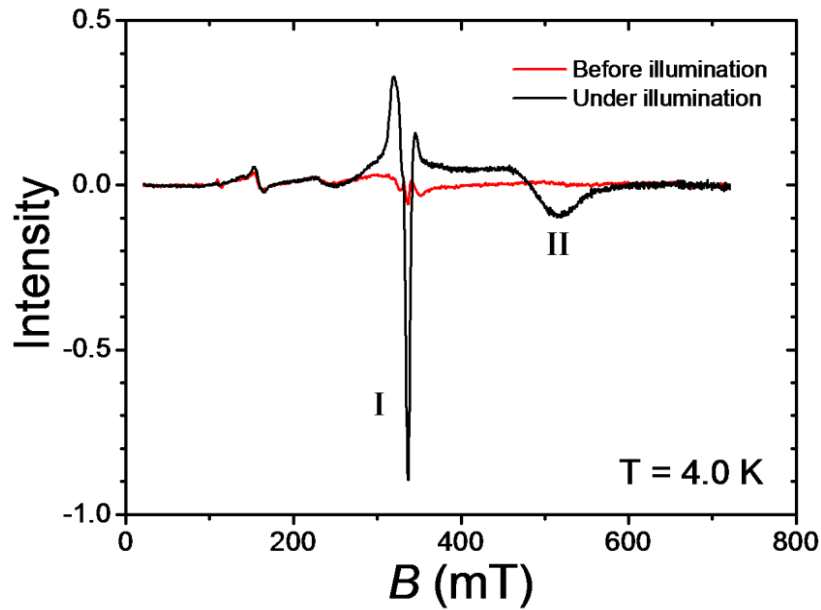

FIG. 1. EPR spectra of the KTN-1.2 crystal at $T = 4.0$ K before UV-light illumination and under a stationary illumination. $B_0$ forms an angle of $\theta \sim 40°$ with the $C_4$ axis in the $C_4$-$C_2$-$C_4$ plane; $B_1 \parallel C_4$.

In Fig. 2, the angular dependence of the photo-EPR spectrum is shown with $B_0$ lying in the $C_4$-$C_2$-$C_4$ crystal plane. Signal I does not reveal any substantial angular dependence. On the contrary, Signal II demonstrates very strong angular dependence of both the resonance field and the signal width. Such spectrum was not previously observed in either KTO or KTN crystals.

The optimal conditions for Signal I and Signal II observations are different. The spectra presented in Figs. 1 and 2 have been recorded with the highest achievable value of the $B_0$ modulation ($\Delta B_0 \sim 2.4$ mT), which is suitable for the wide Signal II. However, the narrower Signal I may be severely distorted with such $\Delta B_0$ value.

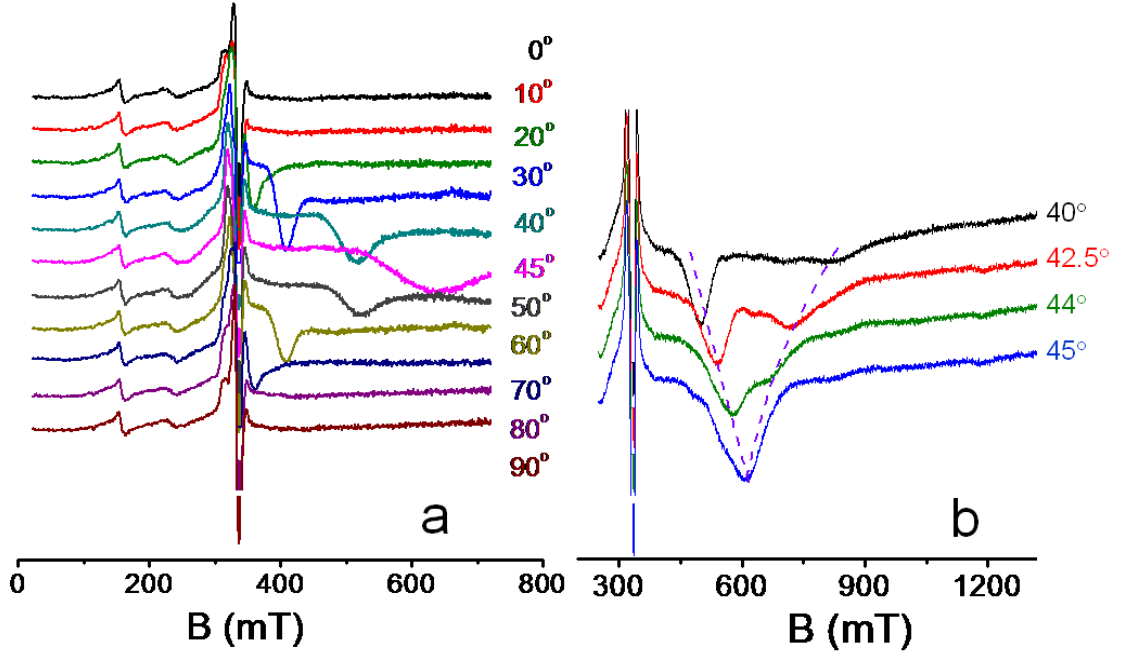

FIG. 2. Angular dependence of the photo-EPR spectrum at $T = 4.0$ K with $B_0$ in $C_4$-$C_2$-$C_4$ plane; $\theta = 0°$ corresponds to $B_0 \parallel C_4$ (a). In panel (b), the two magnetically-nonequivalent centers are observed simultaneously at $\theta \sim 45°$.

Fig. 3 shows the high resolution record of Signal I, the rich structure of which is best revealed when $\theta = 0°$ ($B_0 \parallel C_4$). Measurement of the angular dependence of Signal I has shown that it is caused by several, at least three, photoinduced paramagnetic defects. Its reach structure rapidly smears out with the departure of $\theta$ from 0°. In our opinion, this complex signal can tentatively be assigned to the single trapped charge carriers with $S = ½$. Its structure is caused by the hyperfine interactions between the carrier spin and the nuclear spins of the host lattice ions. It should be mentioned that, in KTN, only oxygen ions do not possess the nuclear magnetic moment; $^{181}$Ta (natural abundance 100%) has a nuclear spin $I = 7/2$, $^{39}$K and $^{41}$K (in total 94%) both have $I = 3/2$, and $^{93}$Nb (100%) has $I = 9/2$. However, we failed to indentify the observed structure using only stationary EPR data. Additionally, the interpretation of the spectrum is complicated because Signals I and II overlap when $\theta = 0°$. In such case, additional complementary methods, such as pulsed EPR, high-frequency EPR and ENDOR spectroscopy, are needed to interpret the photo-EPR signal near $g_{eff} \approx 2$.

At the same time, the data in Fig. 2 reveal a set of specific properties of Signal II: *i*) the resonance field of this component has a strong angular dependence: the resonance field is minimal with $B_0 \parallel C_4$ and rises steeply when $\theta$ deviates from 0°, *ii*) the spectra of this component have a periodicity of 90°; at $\theta$ close to 45°, the two magnetically nonequivalent axial centers oriented along the $C_4$ crystalline axes are revealed; at $\theta = 45°$ ($B_0 \parallel C_2$) these centers are equivalent, *iii*) the width of Signal II is minimal at $\theta = 0°$ and increases severely as $\theta$ increases and *iv*) the lineshape of Signal II is rather asymmetric. It is worth mentioning that as the three $C_4$ axes in the cubic perovskite are equivalent, the actual number of the magnetically nonequivalent centers is three, each with an axis along the $C_4$ axis of the crystal. The third center is not observed in the accessible magnetic field range of the spectrometer.

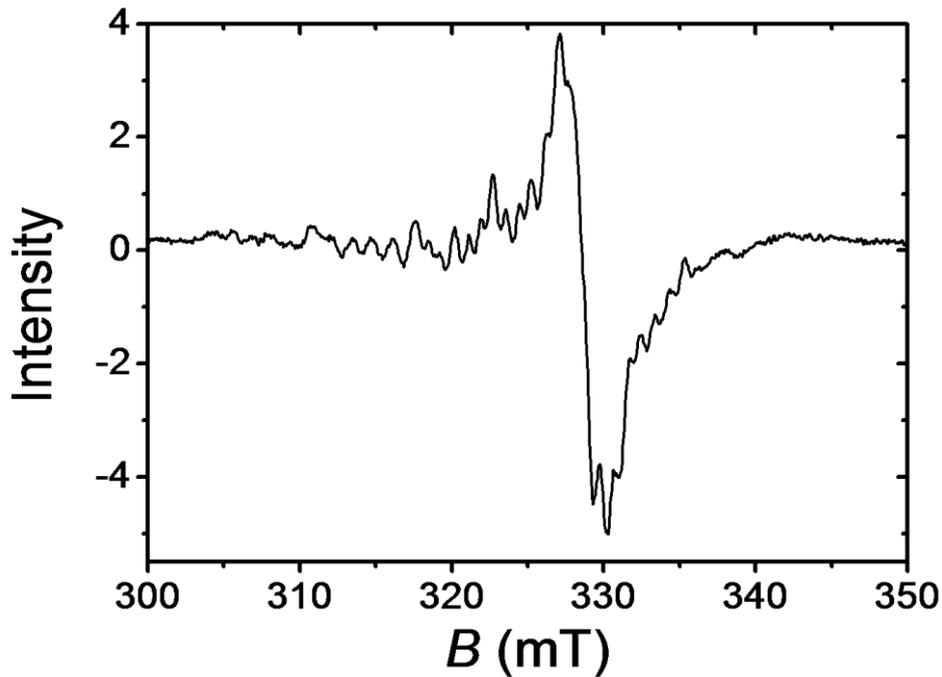

FIG. 3. EPR spectrum of the KTN-1.2 crystal at $T = 4.0$ K, $\Delta B_0 \sim 3$ Oe, $B_0 \parallel C_4$.

Fig. 4a presents the angular dependence of the resonance field for Signal II, where the field values correspond to zero crossing by the signals. The resonance field can also be determined by its correlation with the expressed minimum in the data but such approach leads just to a slight modification of the angular dependence.

As opposed to the photo-EPR spectra of the KTO crystals, the observed angular dependence of the resonance field for Signal II cannot be assigned to the transitions within the isolated spin doublet. The limit case of $g_{\perp eff} = 0$ when the resonance field extends to infinity ($B_{res} = B^0/\cos \theta$) is shown in Fig. 4a by the dashed line. Deviation from the experimental data is observed at $\theta \geq 20°$ and exceeds the uncertainty in the resonance field determination.

Thus, our observations require a new model. Because our objects of interest are the photoinduced centers, it is reasonable to study a more sophisticated than the $S = ½$ center model – the center with $S = 1$. The Hamiltonian of an axial center with $S = 1$ is given by [13]:

$$\hat{H} = D\{\hat{S}_z^2 - \tfrac{1}{3}S(S+1)\} + g_\parallel \beta H_z \hat{S}_z + g_\perp \beta \left(H_x \hat{S}_x + H_y \hat{S}_y\right). \quad (1)$$

In this model, the resonance field exhibits an angular dependence similar to the observed one at the transition between the state with $M_S = 0$ and the closest to it state from $M_S = \pm 1$ doublet if a zero-field splitting $D$ is slightly larger than the microwave quantum (the transition is shown by the arrow in Fig. 4b). The fit of the experimental data is shown in Fig. 4a with the solid line. The magnitudes of the parameters obtained from the fit are $g_\parallel = 0.82 \pm 0.04$, $g_\perp = 0.52 \pm 0.04$ and $D = 0.44 \pm 0.03$ см$^{-1}$.

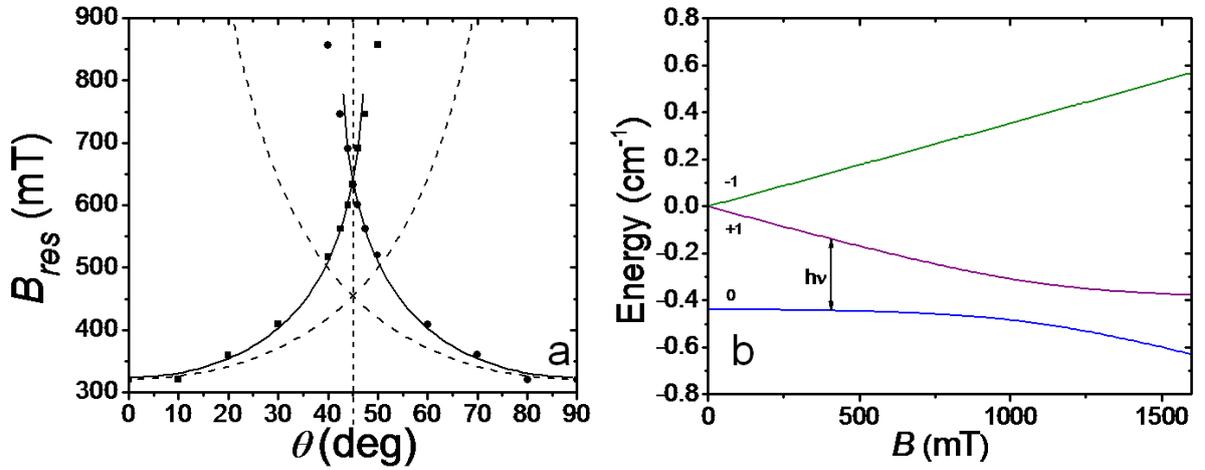

FIG. 4. (a): Angular dependence of the resonance field $B_{res}(\theta)$ for the photo-EPR Signal II. The dashed curves are the angular dependencies $B_{res} = B_{res}^0 / \cos\theta$ with $B_{res}^0 = 312.4$ mT. The solid lines indicate the angular dependence fits with the parameters given in the text. (b): Energy level scheme for the center with $S = 1$ ($\theta = 25°$).

Moreover, the proposed model explains a pronounced broadening of Signal II as $\theta$ deviates from 0°. Two lower levels in Fig. 4b in some range of magnetic field values (~ 1100 mT in the figure) are almost parallel to each other due to mutual repulsion, which is proportional to the off-diagonal matrix elements of the Zeeman interaction. If some distribution in the $D$ values is present, e.g., due to random strains, the spectrum width would be inversely proportional to $d\Delta E/dB$ at $B = B_{res}$, where $\Delta E$ is the energy gap between the levels. Thus, the closer resonance conditions to the field range with minimal $d\Delta E/dB$, which in our case corresponds to an increase in $\theta$, implies a larger spectrum width.

Fig. 5 presents the temperature evolution of the KTN-1.2 photo-EPR spectrum. The intensity of the signal as a function of temperature is shown in Fig. 6 with squares. Under *cw* illumination, Signal II becomes unobservable at T ≥ 10 K (Fig. 5a).

However, if we block the illumination at 10 K and cool the specimen to 4 K in the dark, the total intensity of the signal is recovered. This observation indicates that the paramagnetic centers responsible for Signal II are not destroyed at ~ 10 K. The signal then vanishes due to the center intrinsic properties.

To study the temperature stability of the photoinduced centers, we have used a different approach. The required sample temperature was first reached without illumination. Then, the illumination was switched on for 10 minutes to approach steady-state conditions. Afterwards, the sample was cooled rapidly (within 3-5 seconds) in the dark down to 4.2 K and the spectrum was registered. Prior to these measurements, we have verified that the lifetime of the centers at temperatures below 10 K is much longer than is necessary for the specimen's steady state temperature establishment and spectrum registration. The resulting spectra are shown in Fig. 5b.

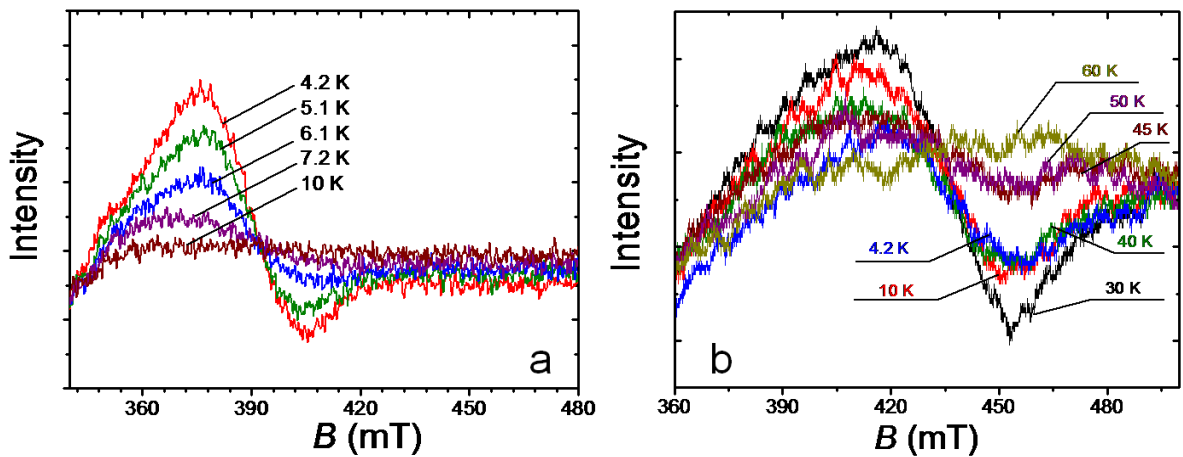

FIG. 5. (a) - Temperature dependence of Signal II in the photo-EPR spectrum of KTN-1.2 at $\theta = 30°$; (b) - photo-EPR spectra observed after blocking the illumination at the temperatures shown in the graph and fast cooling the sample in the dark down to $T = 4.2$ K.

The photo-EPR intensity was evaluated as a difference between the signal magnitudes for a given initial temperature and the initial temperature of 60 K at the signal maximum. The circles in Fig. 6 represent the temperature dependence of the photo-EPR intensity determined by the described approach.

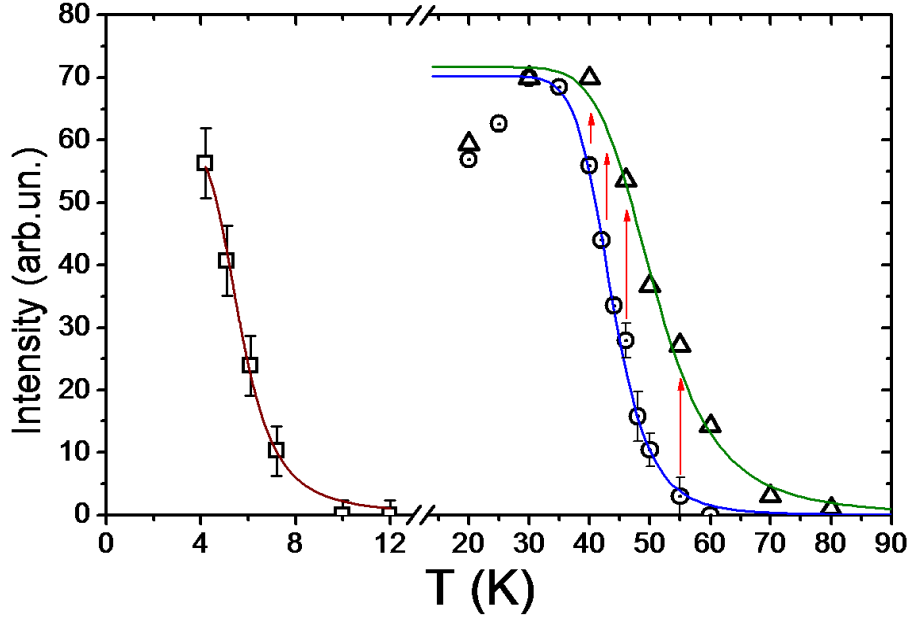

FIG. 6. Temperature dependencies of the photo-EPR signal intensity under *cw* illumination (squares) and after cooling from the given temperature to $T = 4.2$ K with the illumination switched off (circles, see text). Intensity of the photoinduced IR-absorption at ~0.7 eV is shown with triangles [6]. Solid curves are the fits of the data to Eq. 2.

Both of the obtained temperature dependencies of the photo-EPR signal in KTN-1.2 have been fit with the activation model of the center decay using the following expression:

$$I(T) = \frac{A}{1 + B \cdot \exp\left(-E_a / k_B T\right)}, \quad (2)$$

where $E_a$ is the activation energy, $A$ is the amplitude factor, $B$ is the probability ratio of the relaxations via activation mechanism and spontaneous transitions at $T \to \infty$. The fit results are shown in Fig. 6 with solid lines. Both the low temperature dependence and the falling part of the high temperature dependencies are fit well with the used approach. The fit parameter values are $E_a = 3.7 \pm 0.5$ meV, $B = 1900 \pm 1700$ and $E_a = 52 \pm 4$ meV, $B = (93 \pm 9) \cdot 10^5$ for the low and high temperature dependencies, respectively.

## DISCUSSION

Our previous studies have shown that the photoinduced near-IR optical absorption spectra of low concentrated KTN crystals are different from those observed in KTO [6, 7]. The photo-EPR Signal II in KTN-1.2 also has different properties from the anisotropic signals reported earlier for KTO [14, 15]. To determine whether the photoinduced optical spectra and the EPR spectra of KTN-1.2 are related to each other, we compared their temperature dependencies (Fig. 6). One finds immediately that the dependencies of the photo-EPR and the optical spectrum intensities are very similar at high temperature and essentially coincide at temperatures below 40 K. In the 40–70 K range, a difference in the signal intensities develops and increases with the temperature rise. There is a clear reason for such difference. While the optical spectra were recorded in the true steady-state conditions, the photoinduced centers of the photo-EPR case had partially decayed during the 3-5 seconds of sample cooling after switching off the UV-illumination. Taking into account that the lifetime of the photoinduced optical absorption varies from ~100 sec at 1.5 K to a few seconds at 60 K with a steep lifetime shortening when $T > 40$ K, one can realize that the respective discrepancy in the signal intensities should be similar to the observed one. Modification of the temperature dependence of Signal II with an account of the center decay is qualitatively shown in Fig. 6 by the arrows. We believe that the observed dependencies of the photoinduced EPR and IR absorption intensities imply that these signals have the same origin.

The difference between the photoinduced optical absorption and photo-EPR spectra of Signal II in KTN and those found in KTO indicates that the photoinduced paramagnetic centers observed in KTN-1.2 somehow involve the "impurity" Nb ions. The photoinduced optical absorption band of KTN at ~0.7 eV has indeed been assigned to the pair Nb – Nb polarons [7]. Because no photo-EPR spectra typically found in KTO [14, 15] are present in KTN-1.2, the Nb-related centers appear to be the preferable traps for the charge carriers in comparison to the traps present in a nominally pure KTO.

To identify the photo-EPR Signal II of KTN-1.2, we recall that *i*) it originates from the axial centers oriented along the $C_4$ crystal axes, *ii*) it reveals the unusual angular dependence of the resonance field that cannot be fit within an isolated spin-doublet model but can well be described with the $S = 1$ center and *iii*) it is related to the Nb ions.

One can propose just a few models that would satisfy this set of observations. Both $Ta^{5+}$ and $Nb^{5+}$ ions have practically the same ionic radius 0.64 Å in a six-fold octahedral coordination [16]. Thus, minor $Nb^{5+}$ admixture perturbs the crystal lattice of KTO *locally* to the minimum extent. Also, the actual centers cannot be either electron localized on the tantalum ($Ta^{4+}$) or hole on the oxygen ($O^-$) because these should reveal the same properties as the photoinduced centers in KTO.

Thus, the center responsible for the KTN-1.2 anisotropic photo-EPR spectrum could be the $Nb^{4+}$ center with one *d*-electron at the outer shell, whose axial symmetry may be caused by the static Jahn-Teller effect [13, 17]. It is known that the EPR properties of the $d^1$-ions are nontrivial due to the strong influence of the spin-orbit coupling and the dynamic Jahn-Teller effect [13, 18]. Such centers nevertheless should possess a hyperfine structure ($I = 9/2$), which has not been observed. Another argument against the $Nb^{4+}$ center is a high value of the spin-orbit coupling, which is about 750 cm$^{-1}$ for a free ion [19]. Even with a possibility of significant spin-orbit coupling reduction in the crystal and due to the Jahn-Teller effect [20], a remnant value would be enough to produce a well-isolated spin doublet as a ground state. This situation obviously does not fit the condition (*ii*) above. Moreover, we found the spectra of the $Ta^{4+}$ polaron with a resolved hyperfine structure and $g \sim 2$ in the Q-band photoinduced EPR of $KTaO_3$, which can be described as the $S = ½$ centers with a weak *g*-factor anisotropy [21]. We expect the properties of $Nb^{4+}$ centers to be very similar.

The $Nb^{3+}$ centers with the two captured electrons have to be excluded as well. On one hand, these are $S = 1$ particles with an orbital triplet in the ground state that can possess axial symmetry in the case of the static Jahn-Teller effect. On the other hand, such centers studied by EPR did not reveal any unusual *g*-factor values like what we have [22].

The $S = 1$ photoinduced centers in KTN with the axis oriented along the $C_4$ pseudocubic direction of the host crystal related to the niobium may be the $Nb^{4+}$-$Nb^{4+}$ pairs suggested in [6, 7], the $Nb^{4+}$-$Ta^{4+}$ pairs or the $Nb^{4+}$-$O^-$ exciton. The former two are $d^1$-$d^1$ pairs. Following the empiric Goodenough-Kanamori rules [23, 24], the two *d*-electrons should have antiparallel spins, resulting in a nonmagnetic $S = 0$ ground state of the pair. Indeed, all the so far performed experimental studies have shown that such bipolaron-like complexes are EPR silent [25,26].

Thus, the $Nb^{4+}$-$O^-$ exciton in the triplet state is the only appropriate anisotropic photoinduced center responsible for the photo-EPR Signal II in KTN-1.2.

Such a center can manifest all of the EPR spectra details found in our experiments: (*i*) in a fixed configuration, its axis coincides with the $C_4$ crystalline axis; (*ii*) the oxygen *p*- and the niobium *d*-orbital directly overlap and the Goodenough-Kanamori rules [23,24] do not inhibit the ferromagnetic alignment of the spins forming thus the spin triplet; and (*iii*) it involves the Nb impurity.

The $ABO_3$ perovskite-like oxides are highly polarizable materials with the pronounced electron-phonon interaction. Therefore, both the localized *d*-electron on the niobium and the *p*-hole on the oxygen should experience strong electron-lattice interaction, which manifests itself via the formation of polarons and/or the Jahn-Teller effect. The complex thus possesses a set of degrees of freedom that can be revealed in the experiments. First, we believe that both the electron and the hole form polarons. Second, such $Nb^{4+}$-$O^-$ polaronic exciton suggests a possibility of the oxygen hole dynamics: the hole localization at any of the six oxygen ions surrounding the niobium is energetically equivalent. Thus, the $Nb^{4+}$-$O^-$ exciton possesses six energy minima related to the hole localization with a possibility of tunneling through and hopping over the barriers. Similar to the *static* Jahn-Teller effect, the system may be trapped in one of the minima if it is energetically favorable, for example, due to random internal strains and fields. When the hole hopping occurs on a time scale longer than the characteristic time of the experiment (~$10^{-10}$ sec for the X-band EPR), the anisotropic centers will be observed. Otherwise, the averaging within possible axial configurations will smear the spectrum over a wide range of a few thousand Oersted and make it unobservable. We believe that the gradual disappearance of the anisotropic photo-EPR spectrum in KTN-1.2 when $T < 10$ K is related to the activation of the oxygen hole dynamics and that $E_{a1} = 3.7$ meV reflects the average value of the stabilization energy for a given center configuration. It should also be mentioned that the $Nb^{4+}$-$O^-$ exciton is immobile as it is trapped at the Nb impurity.

Third, both $Nb^{4+}(4d^1)$ and $O^-$ ($2p^5$) have orbitally degenerate ground states and are subject to the Jahn-Teller effect. Moreover, the distortions of the nearest surroundings of both centers are coupled to each other. The dynamic Jahn-Teller effect can drastically modify the *g*-factor values of such a center but the theoretical estimate for such sophisticated object is still a challenge.

The $Nb^{4+}$-$O^-$ exciton possesses a huge electric dipole moment but zero charge. The formation of these excitons should lead to the drop of the photoconductivity which was indeed observed in slightly niobium-doped KTO with respect to the undoped one [6]. The presence of the large dipole moment also makes the excitons sensitive to the local electric field and may even be the leading term in making the definite potential minima preferable. The EPR spectra become unobservable at T = 10 K, close to the the FE PT temperature ($T_C$ = 16 K), which may indicate that the stabilization of a definite configuration occurs due to the interaction of the center dipole moment with an internal electric field. In this case, the stabilization becomes less effective as the temperature increases and approaches $T_C$. Then, it becomes clear why the photo-EPR spectrum vanishes in intensity at $T < T_C$.

At first glance, it is surprising that the two centers manifested in the photo-EPR Signal II in Fig. 2 have the same intensities while KTN-1.2 obeys cubic-trigonal FE PT at ~16 K [8]. However, as the low-symmetry phase is distorted along the trigonal axis of the crystal, the tetragonal symmetry centers are left equivalent under such conditions. The center equivalency may also occur due to its relation to the short-range ordered dipole glass phase [27] or the inhomogeneous structure of the polar state of KTN-1.2.

Another observation worth mentioning is the temperature evolution similarity between the photo-EPR Signal II in KTN-1.2 and the anisotropic signals in KTO [14]. In both crystals, the signals vanish at ~10 K under cw illumination, while the centers survive up to higher temperatures, e.g., close to 60 K for KTN-1.2. The spectra of KTO and KTN also reveal the tetragonal symmetry of the photoinduced centers, which probably indicates a similar origin of the photoinduced centers in the KTO and KTN crystals. In KTO, these centers definitely cannot be the $Nb^{4+}$-$O^-$ excitons but can be the $Ta^{4+}$-$O^-$ excitons trapped at various crystal defects. Our preliminary study of the photo-EPR spectra of KTO has revealed the occurrence of another spin-dipole transition in the Q-band (36 GHz) for all tetragonal centers found earlier in KTO in the X-band (10 GHz). Such an observation is incompatible with the models proposed earlier [14, 15].

Interestingly, the $Nb^{4+}$-$O^-$ exciton has the same "$d$-electron + $p$-hole" electronic structure as the so-called "charge-transfer vibronic excitons" $Ta^{4+}$-$O^-$ or $Ti^{3+}$-$O^-$ previously proposed and discussed in [28-30], which act as the source of the intense photoinduced optical absorption and the "green" luminescence in KTO and STO crystals. In particular, the spin-triplet has been predicted to be the ground state of such metastable complexes [29], which is found for the $Nb^{4+}$-$O^-$ excitons in KTN-1.2. These theoretically predicted objects to our knowledge have not been reliably recognized in previous experiments; thus, our results are probably the first experimental detection of these complex species.

## CONCLUSIONS

In summary, the most important result of the paper is recognizing the experimental manifestations of the theoretically predicted polaronic excitons in the diluted $KTa_{1-x}Nb_xO_3$ crystals (x = 0.012). The photoinduced EPR spectrum of KTN-1.2 contains the anisotropic component (Signal II) that originates from the axial centers with $S = 1$, $g_\parallel = 0.82 \pm 0.04$, $g_\perp = 0.52 \pm 0.04$, and $D = 0.44 \pm 0.03$ см$^{-1}$. Our analysis shows that these centers are the polaronic excitons $Nb^{4+}$-$O^-$. The temperature dependence of Signal II is characterized by the two activation energies: $E_{a1} = 3.7\pm0.5$ meV for the internal dynamics that makes the center unobservable above 10 K, and $E_{a2} = 52\pm4$ meV for the center destruction. The comparison of the anisotropic photo-EPR spectrum and the photoinduced optical absorption temperature dependencies indicates that the $Nb^{4+}$-$O^-$ polaronic excitons manifest themselves also via the ~0.7 eV wide absorption band, which arises under UV illumination in the $KTa_{1-x}Nb_xO_3$ crystals with low niobium concentration.

## ACKNOWLEDGEMENTS

This work has been supported in part by the Grants 1M06002 of the MSMT CR, AV ČR AV0Z10100522, SC 02.740.11.5162 and P-RAS "Quantum Physics of Condensed Matter".